\RequirePackage{fix-cm}
\documentclass[twocolumn]{svjour3}          
\smartqed  
\usepackage{graphicx}
\usepackage{mathptmx}      
\usepackage[numbers, sort]{natbib}
\usepackage{mathtools}
\usepackage{amsmath,mathptmx,amssymb,bm}
\usepackage{here}
\usepackage{multirow}
\usepackage{xcolor}
\usepackage{url}
\begin{document}

\title{Empirical Validation of Continuum Traffic Flow Model of Capacity Drop at Sag and Tunnel Bottlenecks}
\subtitle{}


\author{Shin-ichiro Kai \and Ryota Horiguchi \and Jian Xing \and Kentaro Wada
}


\institute{S. Kai \at
              i-Transport Lab. Co., Ltd., 
              Chiyoda-ku, Tokyo, 101-0052, Japan.\\
              \email{kai@i-transportlab.jp}           
           \and
           R. Horiguchi \at
              i-Transport Lab. Co., Ltd.,  
              Chiyoda-ku, Tokyo,  101-0052, Japan.\\
              \email{rhoriguchi@i-transportlab.jp}
           \and
           J. Xing \at
             Nippon Expressway Research Institute Co., Ltd.,  
             Machida, Tokyo, 194-8508, Japan.\\
              \email{xing@ri-nexco.co.jp} 
           \and
           K. Wada \at
             Institute of Systems and Information Engineering, University of Tsukuba, 
             Tsukuba, Ibaraki, 305-8573, Japan.\\
              \email{wadaken@sk.tsukuba.ac.jp} (Corresponding Author)             
}

\date{Received: date / Accepted: date}

\maketitle

\begin{abstract}
This study validates the continuum traffic flow model of capacity drop at sag and tunnel bottlenecks, as proposed by \citet{Jin2018} and \citet{Wadaetal2020}, through empirical analysis.
Specifically, {after addressing the limitations in the existing studies}, we calibrate the model using data from multiple congestion events at several expressway bottlenecks.
{We then demonstrate that the model can reproduce the observed speed recovery near the head of the queue, and assess whether both estimated bottleneck capacities and locations are consistent with observed traffic conditions.}
{Finally, as an application of the calibration results}, we examine the relationship between the spatial changes in the estimated traffic capacity and longitudinal gradients.

\end{abstract}

\section{Introduction}
\label{Ch1.1}
At bottlenecks such as sags, tunnels, merges, and lane-drop on expressways, it has been observed that the traffic flow rate during congestion, i.e., queue discharge flow rate (QDF), decreases by approximately 10\% compared to the rate immediately after the onset of congestion.
This phenomenon is referred to as capacity drop (CD) \citep{Wada2021a} and has been studied for many years \citep[e.g.][]{Edie1958, Banks1991, Hall1991, Cassidy1999}.
The occurrence of the CD phenomenon increases the length of congestion queues, resulting in increased travel time.

With respect to the CD phenomenon in sag and tunnel bottlenecks, \citet{Koshi1986} hypothesizes that it occurs due to the extremely low acceleration rate of vehicles at the head of the queue.  
\citet{Koshi1986} also suggests that the QDF is characterized by this low acceleration.
Based on findings from a series of studies by Koshi et al. \citep{Koshi1986, Koshi1992, Koshi1993}, various countermeasures have been implemented on Japanese expressways to increase QDF by improving the acceleration of vehicles at the head of the queue.
For example, attempts have been made to provide information about the location where the congestion ends through LED signs \citep{Yamada2003, Nakatani2005} or speaker announcements \citep{Sato2021, Wada2023}.
In addition, installing moving light guides designed to induce vection has also been implemented \citep{Kameoka2013, Shiomi2017, Masumoto2018}.
Although these countermeasures have demonstrated certain effectiveness, the mechanisms behind the CD phenomenon remain unclear, so they have been implemented in a trial-and-error manner.

Recently, the continuum traffic flow model of the CD phenomenon at sags and tunnels on expressways has been proposed by \citet{Jin2018} and \citet{Wadaetal2020}. 
\citet{Jin2018} presents a behavioral kinematic wave model to explain the bottleneck effects at sag and tunnel bottlenecks. 
Assuming increasing time gaps, Jin derives location-dependent (i.e., inhomogeneous) triangular fundamental diagrams (FDs) and imposes a bounded acceleration (BA) constraint on the stationary states inside the capacity reduction zone to account for the CD phenomenon. 
\citet{Wadaetal2020} extended the model of \citet{Jin2018} to a second-order formulation that captures both the formation and temporal evolution of the capacity drop phenomenon. 
By examining the model properties, they revealed why and how the capacity drop occurs as a result of bottleneck effects and BA. 

As an application of this theory, \citet{Wada2022} estimated bottleneck characteristics such as its location and traffic capacity at a sag bottleneck on an expressway by calibrating the model with empirical data.
However, the general applicability of this approach remains to be validated, as the study focused on a single site and specific congestion events.
To address this issue, \citet{Kai2023} calibrates the model at several sites using the method of  \citet{Wada2022}. 
Nevertheless, {as discussed in detail in the next section, this study still has the following issues: (i) the selection of target congestion events; (ii) the preprocessing of probe data; and (iii) unnecessary constraints on parameters.
Furthermore, neither \citet{Wada2022} nor \citet{Kai2023} examined the validity of the estimated bottleneck characteristics in terms of both capacity and location.}

{This study validates the continuum traffic flow model of capacity drop at sag and tunnel bottlenecks, as proposed by \citet{Jin2018} and \citet{Wadaetal2020}, through empirical analysis.
Specifically, after addressing the limitations outlined above in existing studies, we calibrate the model using data from multiple congestion events at several expressway bottlenecks.
We then demonstrate that the model can reproduce the observed speed recovery near the head of the queue, and assess whether estimated bottleneck capacities and locations are consistent with observed traffic conditions.}
{Finally, as an application of the calibration results,} we examine the relationship between the spatial changes in the estimated traffic capacity and longitudinal gradients.

The remainder of this paper is organized as follows.
{
Section \ref{Ch1.2} reviews the existing literature.
Section \ref{Ch2} presents an overview of the continuum traffic flow model at sag and tunnel bottlenecks. 
Section \ref{Ch3} describes the calibration and validation frameworks.
Section \ref{Ch4} presents the results of the empirical analysis, including both the calibration results and their validation.
Finally, Section \ref{Ch6} concludes the paper.
}

\section{Literature Review}
\label{Ch1.2}
The occurrence of the CD phenomenon is attributed to driving behavior.
The hypotheses proposed to explain this phenomenon can mainly be divided into two categories \citep{Wada2021a}: one is (longitudinal) sluggish car-following behavior, and the other is (lateral) disruptive lane-changing behavior. 
Regarding the latter, \citet{Laval2006} hypothesize that lane-changing vehicles create voids in the traffic flow, resulting in a decrease in flow rate. However, since CD phenomena have been reported even at sags or tunnels on single-lane roads \citep[e.g.,][]{Okamura2000, Yoshikawa2005}, these hypotheses are neither necessary nor sufficient to explain the occurrence of CD.

With respect to car-following behavior, it is generally thought that a reduction in the QDF is caused by low acceleration and/or delayed responses of vehicles departing from the queue, or by traffic disturbances (e.g., oscillations) upstream of bottlenecks \citep{Hall1991, Koshi1992}.
Based on this idea, the CD phenomenon has been modeled using increased driver reaction times \citep{Zhan2005} and by considering drivers who become less aggressive and adopt longer response times and minimum spacing when passing traffic disturbances \citep{Chen2014}.
However, this approach essentially imposes CD exogenously; furthermore, although bottlenecks are likely to induce sluggish behavior, the relationship between such behavior and CD remains unclear.

In contrast, the continuum traffic flow model proposed by \citet{Jin2018} and \citet{Wadaetal2020} provides an endogenous description of the CD phenomenon at sag and tunnel bottlenecks on expressways.
As mentioned above, this model is able to explain the relationship between CD and bottlenecks. 
Using this feature of the model, {\citet{Wada2022} and \citet{Kai2023} conducted model calibration. However, as noted earlier, three limitations (i)--(iii) in the studies may adversely affect calibration results:
\begin{itemize}
	\item[(i)] Some events used in previous studies were affected by downstream congestion, meaning that the bottleneck was no longer active. In such cases, the calibration results cannot be properly interpreted. 
	\item[(ii)] The probe data used are sampled at relatively long intervals, yet no appropriate preprocessing (such as smoothing) was applied. As a result, the average speed in the recovery profiles changed abruptly, which likely affected the estimation of the acceleration-related parameters. 
	\item[(iii)] The model has two free-flow speed parameters: one in the FD and the other in the BA model. Because they play different roles and refer to different locations, treating them as identical, as done in previous studies, may distort the calibration. 
\end{itemize}
Furthermore, previous studies assessed only the validity of the estimated bottleneck capacity and did not examine the validity of the estimated bottleneck location.
This study addresses the above limitations in both calibration and validation and conducts an empirical analysis based on the improved framework, which is a key contribution.}

\begin{figure}[tb]
\centering
\includegraphics[width=8cm]{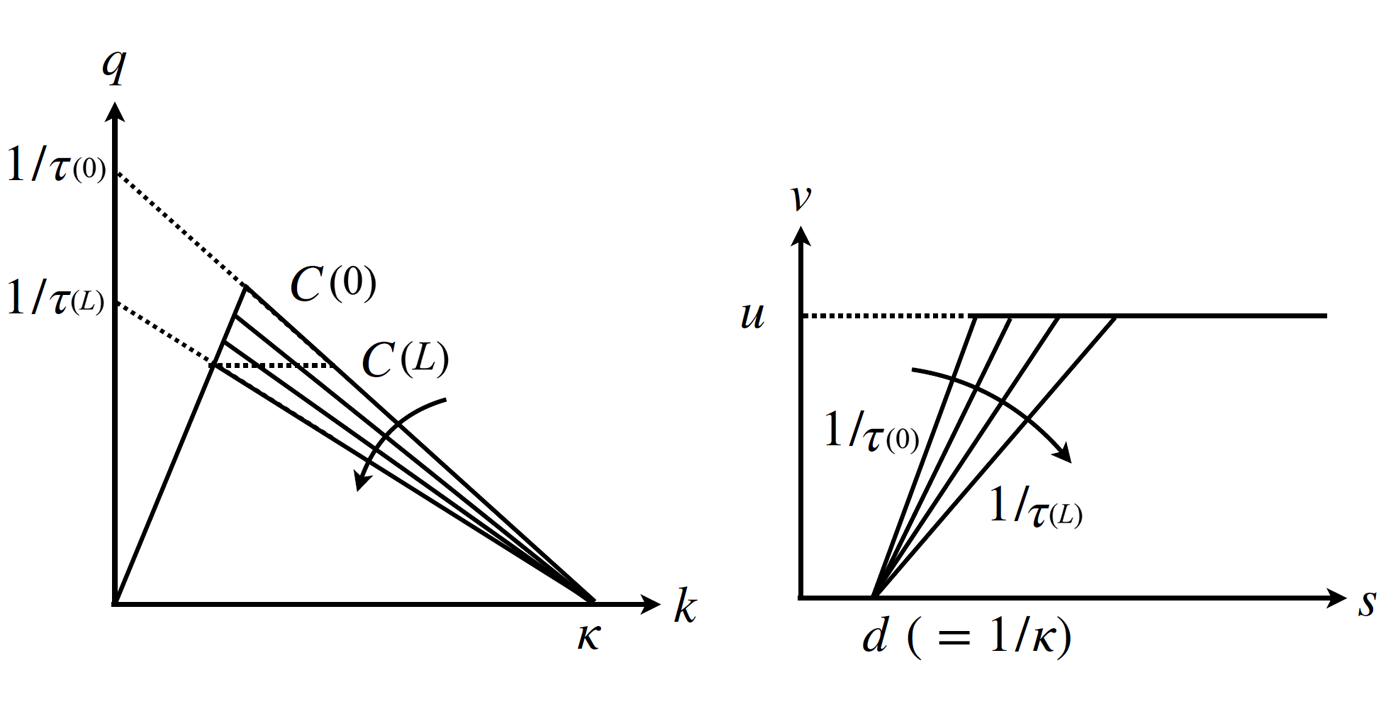}
\caption{Flow-density FD (left) and speed-spacing FD (right)}
\label{fig1}
\end{figure}

\section{Model}
\label{Ch2}

In the continuum traffic flow model proposed by \citet{Jin2018} and \citet{Wadaetal2020}, sag and tunnel bottlenecks are assumed to be {\it sections} where traffic capacity continuously declines in space.
A similar concept is also presented by \citet{Coifman2011}.
However, this approach contrasts with the conventional modeling commonly adopted in the literature, where bottlenecks are explicitly or implicitly treated as {\it points}.

Specifically, the bottlenecks are modeled as an inhomogeneous FD, as illustrated in Figure \ref{fig1}.
The left part of the figure shows the flow-density triangular FD.
Suppose the bottleneck is located between $x = 0$ and $x = L$; then the reduction in capacity $C(x)$ from $C(0)$ to $C(L)$ is modeled by a gradual decrease in the slope of the congested branch of the FD.
This implies that the safe time gap $\tau(x)$ (reciprocal of the y-intercept in the flow-density plane), which refers to the minimum time gap required to maintain a safe distance gap, increases spatially.
The relationship between $C(x)$ and $\tau(x)$ is as follows:
\begin{equation} \label{Eq1}
\quad C(x) = \frac{u\kappa}{1+u\kappa\tau(x)} \qquad 0 \leq x \leq L,  
\end{equation}
where the free-flow speed $u$ and the jam density $\kappa$ are assumed to be constant.
This bottleneck effect is illustrated in the speed–spacing FD shown in the right part of Figure~\ref{fig1}.
It reflects driver behavior in which the reduction in speed in sags or tunnels may occur unconsciously, causing drivers to continue maintaining their distance gap (or spacing).

Another assumption of the model is that the acceleration rate is bounded, reflecting physical limitations such as vehicle mechanics, driver response, and road geometry.
In the context of sag and tunnel bottlenecks, this BA state may represent a situation in which drivers fail to correctly recognize that they have exited the bottleneck section and, as a result, do not sufficiently press the accelerator pedal \citep{Wada2022}.

\begin{figure}[tb]
\centering
\includegraphics[width=8cm]{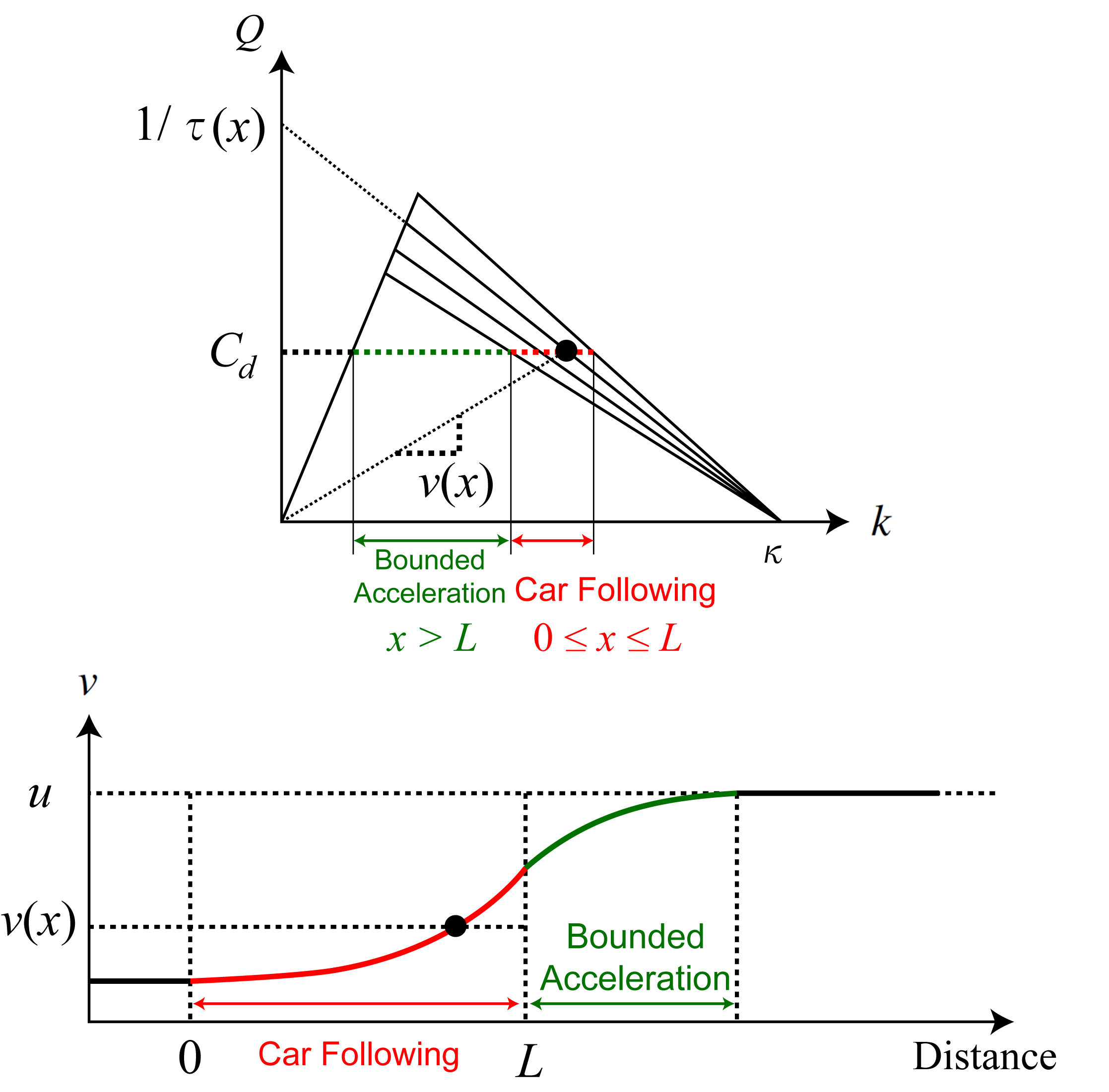}
\caption{Relationship between the FD in the capacity drop stationary state (upper) and the speed recovery profile (lower)}
\label{fig2}
\end{figure}

Under the above two assumptions, let us consider the capacity drop stationary state, which refers to a persistent congested condition in which the flow, speed, and very low acceleration near the bottleneck {section} remain nearly stationary and stable.
In this state, the BA (non-equilibrium) state arises downstream of the bottleneck {section} ($x \geq L$), while the bottleneck section ($0 \leq x \leq L$) follows the car-following (equilibrium) state \citep{Wada2022}, as illustrated in Figure~\ref{fig2}. 
More specifically, when the QDF stabilizes at $C_{d}$, the speed within the bottleneck section is described by the congested branch of the FD, that is, 
\begin{equation} \label{Eq2}
\quad v(x) = \frac{1}{\{1/C_d-\tau(x)\} \kappa} \qquad 0 \leq x \leq L.  
\end{equation}
Given the assumption that the safe time gap satisfies $\tau_{x}(x) = \mathrm{d}\tau(x)/\mathrm{d}x > 0$ within the bottleneck section, the speed $v(x)$ becomes a strict increasing function of $x$.
Furthermore, when $\tau_{xx}(x) = \mathrm{d}^2\tau(x)/\mathrm{d}x^2 > 0$, the shape of the speed recovery profile becomes convex, as typically observed in practice (see the lower part of Figure~\ref{fig2}). 
Note that the acceleration of traffic flow in a stationary state can be expressed as $a(x) = v(x)v_x(x)$ \citep{Jin2018}.
Substituting Equation~\eqref{Eq2} into this expression, we obtain 
\begin{equation} \label{Eq3}
\quad a(x)= \frac{\tau_x(x)}{\{1/C_d-\tau(x)\}^3 \kappa^2} \qquad 0 \leq x \leq L.
\end{equation}

On the other hand, the BA state downstream of the bottleneck {section} is assumed to be described by the TWOPAS model, which is a typical model expressing acceleration limits \citep{Jin2018}: 
\begin{equation} \label{Eq4}
\quad a(x, v)= \{a_0 - g\Phi(x)\}(1 - v(x)/u) \qquad x \geq L,  
\end{equation}
where $a_0$ is the acceleration parameter, $g$ is the gravitational acceleration, and $\Phi(x)$ is the decimal gradient at $x$.
Under this equation, the speed recovery profile (i.e., the solution to $\mathrm{d}v/\mathrm{d}t = a(x, v)$) is concave unless $\Phi_x(x) = \mathrm{d}\Phi(x)/\mathrm{d}x < 0$ (e.g., crest sections).
The BA state is depicted in the interior of the FD (see the upper part of Figure~\ref{fig2}).

According to the model, the head of the queue in a capacity drop stationary state corresponds to the location where the car-following state transitions to BA state, that is, the downstream end $x = L$ of the bottleneck section.
This contrasts with the view of \citet{Koshi1992,Koshi1993}, which regarded the head of the queue as the location where speed recovery begins (i.e., $x=0$ in this model). 
This discrepancy may arise from whether the bottleneck is regarded as a {\it section} or a {\it point}.

Another noteworthy point is the relationship between the bottleneck section and the point where flow breakdown occurs.
When demand exceeding the bottleneck capacity $C(L)$ reaches the bottleneck {\it section}, the point of flow breakdown varies within $0 \leq x \leq L$ depending on the level of demand.
In all cases, however, flow breakdown occurs upstream of the head of the queue in the capacity drop stationary state ($x = L$), and downstream of the location where speed recovery begins ($x = 0$).

Finally, let us look at the relationship between QDF $C_d$ in a capacity drop stationary state and the parameters $\tau(x)$, $\tau_x(x)$, and $a_0$.
Since the QDF is determined at the head of the queue, the relationship between these variables at $x = L$ is important.
Specifically, since the location $x = L$ can be interpreted as either a car-following state or a BA state, the following relationship must be satisfied:
\begin{equation} \label{Eq5}
\quad \frac{\tau_x(L)}{\{1/C_d-\tau(L)\}^3 \kappa^2}  = \{a_0 - g\Phi(L)\}\biggl(1 - \dfrac{v(L)}{u}\biggr).
\end{equation}

\noindent Substituting Equation \eqref{Eq2} into this and simplifying, we obtain, 

\begin{align}  
& \quad \frac{\{a_0 - g\Phi(L)\}}{\tau_x(L)}\left\{\frac{1}{C_d}-\tau(L)\right\}^2  \notag\\
& \mspace{150mu}\left( \left\{ \frac{1}{C_d} - \tau(L)\right\}\kappa^2 - \frac{\kappa}{u}\right) = 1. 
\label{Eq6}  
\end{align}  
From this equation, we can see that $C_d$ and $\tau(L)$ are inversely proportional when other parameters are fixed, as well as that $C_d$ increases as $\tau_x(L)$ decreases and $C_d$ increases as $a_0$ increases.
The first relationship shows that the time gap at the head of queue becomes smaller, and the traffic flow increases as vehicles follow at shorter spacing. 
This reduction in headway also increases the bottleneck capacity $C(L)$ (see also Equation \eqref{Eq1}).
The last relationship shows that the traffic flow improves when vehicles pass the head of the queue with greater acceleration.
To interpret the second relationship,  it is important to understand that $\tau_x(x)$ indicates how smoothly the speed of a following vehicle is adjusted in response to the speed of the lead vehicle.
Specifically, the speed $v_n(t, x)$ of vehicle $n$ at location $x$ at time $t$ in the car-following state, and the speed $v_{n-1}(t-\tau(x), x+d)$ of the lead vehicle, are related as follows (see 
\cite{Wadaetal2020}, for details): 
\begin{equation} \label{Eq7}
\quad v_n(t, x) =  \biggl[\frac{1}{v_{n-1}(t-\tau(x), x+d)} + \tau_x(x)\biggr]^{-1}.
\end{equation}
If $\tau_x(x) = 0$, the following vehicle matches the speed of the lead vehicle while maintaining a safe time gap and minimum spacing.
As $\tau_x(x)$ increases, the following vehicle's speed becomes lower than that of the lead vehicle, indicating a delay in speed adjustment.
Therefore, during speed recovery, a smaller value of $\tau_x(L)$ results in faster adaptation to the lead vehicle's speed, thereby improving traffic flow.

\section{{Calibration and Validation Frameworks}}
\label{Ch3}

{This section provides a detailed description of the frameworks for model calibration and validation.
In this study, calibration refers to determining the model parameters so that the model reproduces the observed speed recovery profiles in the capacity drop stationary state.
However, whether the parameters obtained through this calibration are valid is another issue.}

{A key feature of the model used in this study is that it explains the CD phenomenon as a consequence of the bottleneck influence, which is also the main cause of flow breakdown. This contrasts with previous models that attributed the CD phenomenon to exogenous changes in parameters between free-flow and congested conditions \citep{Zhan2005,Chen2014}. Therefore, it is essential to evaluate whether the estimated bottleneck capacity and location obtained from the calibrated parameters are consistent with the flow breakdown mechanism. This is the basis of the validation used in this study.}

{In the following, we first describe the calibration method and then explain the validation procedure in detail.}

\subsection{{Calibration Method}}
In this study, we basically follow the calibration method proposed by \citet{Wada2022}.
{However, neither \citet{Wada2022} nor \citet{Kai2023} provided a systematic description of the calibration procedure.
Therefore, we present not only an overview of the calibration method but also a detailed explanation of each step.}

The calibration method estimates the bottleneck section and parameters $\tau(x)$ and $a_0$ for each congestion event, based on the QDF $C_d$ and average speed recovery profile $v(x)$ observed in a capacity drop stationary state.
The calibration consists of three steps: (i) setting the bottleneck section $(x = 0, L)$, (ii) estimating the time gap $\tau(x)$, and (iii) estimating the acceleration parameter $a_0$.
Here, $u$ and $\kappa$ are set in advance to appropriate values based on observed data and/or rules of thumb.
As mentioned in {Section \ref{Ch1.2}}, the original method did not fully utilize the flexibility of the model, as it assumed the same free-flow speed within the bottleneck section and downstream of it.
In this study, the model is calibrated by assigning different variables to the free-flow speed:  $u_{FD}$ for the bottleneck section{, which affects the capacity,} and $u_{BA}$ for the BA section{, which determines the final recovery speed downstream of the bottleneck section}, in order to achieve a better fit for each section\footnote{{Although one additional parameter is introduced compared with the previous method, it is set from the observed data. Thus, it does not cause overfitting in the subsequent calibration process.}}.

In step (i), the location where speed recovery begins is first identified from the speed recovery profile, and a nearby point is set as the upstream end $x = 0$ of the bottleneck section.
Next, the downstream end of the bottleneck is set at some location $x > 0$.
In the typical case where the speed profile changes from convex to concave, an appropriate point near the inflection of the profile is chosen as $x = L$.
As described later, the bottleneck section should not be determined solely based on the shape of the speed profile, but should be adjusted to ensure a good fit of the model.

In step (ii), the time gap $\tau(x)$ within the bottleneck section $(0 \leq x \leq L)$ is estimated from the relationship in Equation~\eqref{Eq2}.
{We first calculate the average speed by aggregating the observed speeds of individual probe vehicles over a fixed spatial interval $\Delta x$, and then estimate $\tau(x)$ for each interval using Equation~\eqref{Eq2}. The estimated values of $\tau(x)$, denoted as $\tau(x_1), \tau(x_2), \dots, \tau(x_M)$, are obtained at the locations $x_1 (=0), x_2 (=\Delta x), …, x_M (=L)$ for $0 \leq x \leq L$. From these estimates, we can also approximate $\tau_x(L)$ as $\tau_x(L) \approx \{\tau(L) - \tau(L-\Delta x)\}/\Delta x$. However, this approximation is susceptible to local data variability and measurement error, as it relies on discrete (non-continuous) observations. We thus further fit a continuous, monotonically increasing function $f(x \mid \bm{\beta})$, with parameter vector $\bm{\beta}$, to the estimated safe timegap values. More specifically, a quadratic function $f(x| \bm{\beta}) = \beta_1x^2 + \beta_2x +\beta_3$ is assumed, and the  parameter vector $\bm{\beta}$ is determined by solving the following problem:
	\begin{align}
		\quad \min_{\bm{\beta}}. \sum_{i=1}^M |\tau(x_i)-f(x | \bm{\beta}) | \quad \mathrm{s.t.} \ f(x_M | \bm{\beta}) = \tau(x_M).
	\end{align}
	\noindent The constraint is imposed to ensure consistency between the speed $v(L)$ in the BA model (right-hand side) with the value of $\tau(L)$ in the FD (left-hand side) in Equation~\eqref{Eq5}.}
	{The problem can be solved using the polynomial curve-fitting functions available in standard numerical computation libraries.}

Finally, in step (iii), we estimate the acceleration parameter $a_0$ by substituting the previously estimated parameters and the gradient data $\Phi(L)$ into Equation~\eqref{Eq5}{, and then assess the model fit. Here, the goodness of fit refers to how well the model reproduces the observed speed recovery profile. However, close agreement in the bottleneck section is expected, since $\tau(x)$ is estimated by fitting the model to the observed speed profile. Therefore, the key criterion for evaluating model fitting is how accurately the BA model reproduces the speed profile downstream of the bottleneck section. Although the BA state typically extends for several kilometers until free-flow speed is recovered, the QDF is determined by the traffic flow immediately downstream of the bottleneck. Hence, it is sufficient to assess the model fit within a few hundred meters to one kilometer downstream from $x = L$.}

{There is a remark for finalizing the calibration.} As can be seen from the above procedure, once the bottleneck section is specified, the parameters $\tau(x)$ and $a_0$ are uniquely determined. Thus, selecting the bottleneck section effectively serves as a tuning parameter to improve the model's goodness of fit. For this reason, if necessary, one should return to step (i) to adjust the bottleneck section, after assessing the model fit in step (iii). 

\subsection{{Validation}}

{As mentioned at the beginning of this section, in the validation, we assess whether the estimated capacity and location of the bottleneck section are reasonable in light of the flow breakdown mechanism. This aims to verify the core idea of the model, which explicitly links the bottleneck effect to the CD phenomenon. The observed data used for this validation are not from the persistent congested condition used for calibration, but from just prior to the flow breakdown.}

{
The validation of capacity relies on the mechanism in which flow breakdown occurs when traffic demand exceeds the bottleneck capacity. We here follow the approach of \citet{Wada2022} and \citet{Kai2023}. By substituting $u_{FD}$, $\kappa$, and the estimated $\tau(x)$ into Equation~\eqref{Eq1}, we calculate the bottleneck capacity $C(x)$ along the section. We then compare its minimum value $C(L)$ to the observed traffic flow rate just prior to flow breakdown (i.e., breakdown flow rate (BDF)). As demand gradually increases until breakdown, it is reasonable to expect that the BDF and the bottleneck capacity $C(L)$ take similar values. 
}

{The validation of the bottleneck location is conducted by comparing the estimated section with the area where the speed reduction occurs just prior to the flow breakdown.
As described in Section~\ref{Ch2}, the model expects that the flow breakdown occurs within the bottleneck section, that is, between $x = 0$ and $x = L$. In particular, if demand gradually increases, the greatest speed reduction is expected near $x = L$, where the capacity is at its minimum. 
}

\color{black}
\section{Empirical Analysis}
\label{Ch4}
To validate the continuum traffic flow model, this section presents an empirical analysis of multiple congestion events at several sag bottlenecks on Japanese expressways. First, we explain the target sites. Next, we explain the data required for model calibration {and validation}, along with the data processing. We then present the model calibration and {validation} results. {Finally, we show an application of the calibrated model.}

\subsection{Target Sites}
\label{Ch4.1}
Analysis target sites and congestion events are selected based on congestion event data collected by the Nippon Expressway Companies (NEXCOs).
{To analyze various sag bottlenecks, we include both two-lane and three-lane sections, as well as sections that are predominantly uphill or downhill.} 
Accordingly, four locations where congestion frequently occurs are chosen, as listed in Table \ref{Tab1}. 
At Nisshin and Semimaru, a merging ramp is located upstream of the analysis section. 
At Takasaka, an off-ramp to a rest area (Service Area, SA) is located downstream of the section. 
At Komagawa, a merging ramp is situated downstream. 
{Based on data and field surveys, we confirmed that the sag is the primary cause of flow breakdown and persistent congestion at these sites, though these facilities may also contribute.}

For each section, we selected 10 congestion events. The selected events satisfy the following criteria: (a) the head of the queue is confirmed to be located within the target section; (b) the congestion area fully covers the sag section; and (c) the number of probe samples is sufficient. {Specifically, Criterion (a) is for ensuring that the selected congestion is caused by the sag section itself, rather than by congestion downstream of the sag. Criterion (b) is a prerequisite for the calibration. The method relies on the model property that the upstream end of the bottleneck section corresponds to the point where speed recovery begins during congestion. To apply this property, the tail of the queue must extend upstream of the sag that is assumed to contain the bottleneck section. Criterion (c) is required to obtain a proper speed recovery profile. The probe data used in this study are sampled at relatively long intervals. To derive a continuous and smooth profile from such data, a sufficient number of probe samples must be averaged, along with appropriate smoothing of individual data.}

\subsection{Data and Pre-processing}
\label{Ch4.2}
The data required to calibrate the model described in Section \ref{Ch3} consist of the QDF obtained from detector data, and the speed profile during the capacity drop stationary state acquired using probe data. 
To ensure that the QDF is calculated under sufficiently stable discharge flow conditions (i.e., the capacity drop stationary state), we use the 5-minute average flow rate from 30 minutes after the start of the congestion period to 10 minutes before its end (or from 15 minutes after the start to 10 minutes before its end if the congestion period is less than 1.5 hours).

\begin{table}[tbp]
\caption{Target sites}
\label{Tab1}
{
\tabcolsep = 3pt
\begin{tabular}{lllcc}
\hline
\multicolumn{1}{c}{\begin{tabular}[c]{@{}c@{}}Bottleneck\\name\end{tabular}} & \multicolumn{1}{c}{\begin{tabular}[c]{@{}c@{}}Expressway\\ name\end{tabular}} & \multicolumn{1}{c}{Direction} &  \begin{tabular}[c]{@{}c@{}}Num of lane\\ (by direction)\end{tabular} & \begin{tabular}[c]{@{}c@{}}Longtitude\\{gradient [}\%{]}\end{tabular} \\ \hline
Nisshin & Tomei & {\begin{tabular}[c]{@{}c@{}}east bound\\ \end{tabular}} & 2 & $-$2.7 - 1.6 \\ \hline
Semimaru & Meishin & {\begin{tabular}[c]{@{}c@{}}east bound\\ \end{tabular}} & 2 & $-$3.5 - 3.5 \\ \hline
Takasaka & Kan-etsu & {\begin{tabular}[c]{@{}c@{}}south bound\\ \end{tabular}} & 3 & $-$0.4 - 1.1 \\ \hline
Komagawa & Kan-etsu & {\begin{tabular}[c]{@{}c@{}}north bound\\ \end{tabular}} & 3 & $-$1.2 - 1.0 \\ \hline
\end{tabular}
}
\end{table}

The speed profile is obtained by calculating the average speed every 100 [m] in the vicinity of the target section, using only the times during which the QDF is calculated. These speeds are derived from probe data collected by the Japanese toll collection system known as “ETC2.0.”
This probe data is recorded at intervals of approximately 200 [m], which causes the recorded values to fluctuate at each point.
To address this issue, we apply a symmetric exponential moving average (sEMA) filter \citep{Thiemann2008} to smooth the speed profile.

Other parameters are set as follows.
The parameter $\kappa$ is set at 140 [veh/km/lane].
One of the free-flow speed parameter, $u_{FD}$, is determined based on the free-flow speed of the flow-density plots from the loop detector. Another parameter, $u_{BA}$, is determined from speed profile obtained from probe data in the BA section.

{For validation, the BDF is defined as the 15-minute flow rate just prior to the flow breakdown, calculated from detector data.} {Also, we use the average speed data, calculated from the smoothed probe data aggregated over 15, 10, and 5 minutes before the flow breakdown, to identify where the speed reduction occurs.}

\subsection{Calibration Results}
\label{Ch4.3}
\begin{figure}[tbp]
\centering
\includegraphics[height=5.4cm]{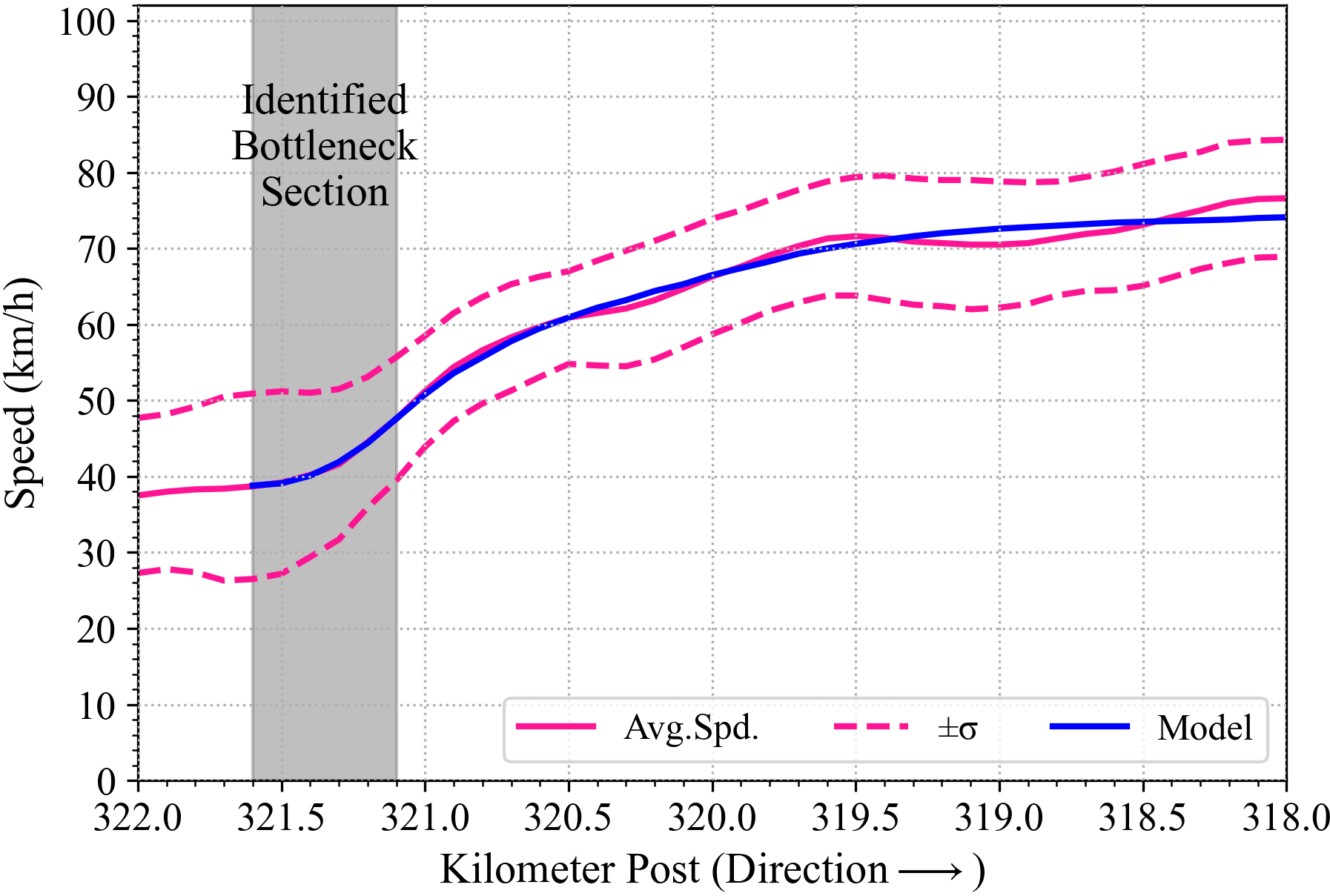}
\caption{Example of model calibration for Nisshin (Sep. 30, 2019)}
\label{Fig4-1}
\end{figure}

\begin{figure}[tbp]
\centering
\includegraphics[height=5.4cm]{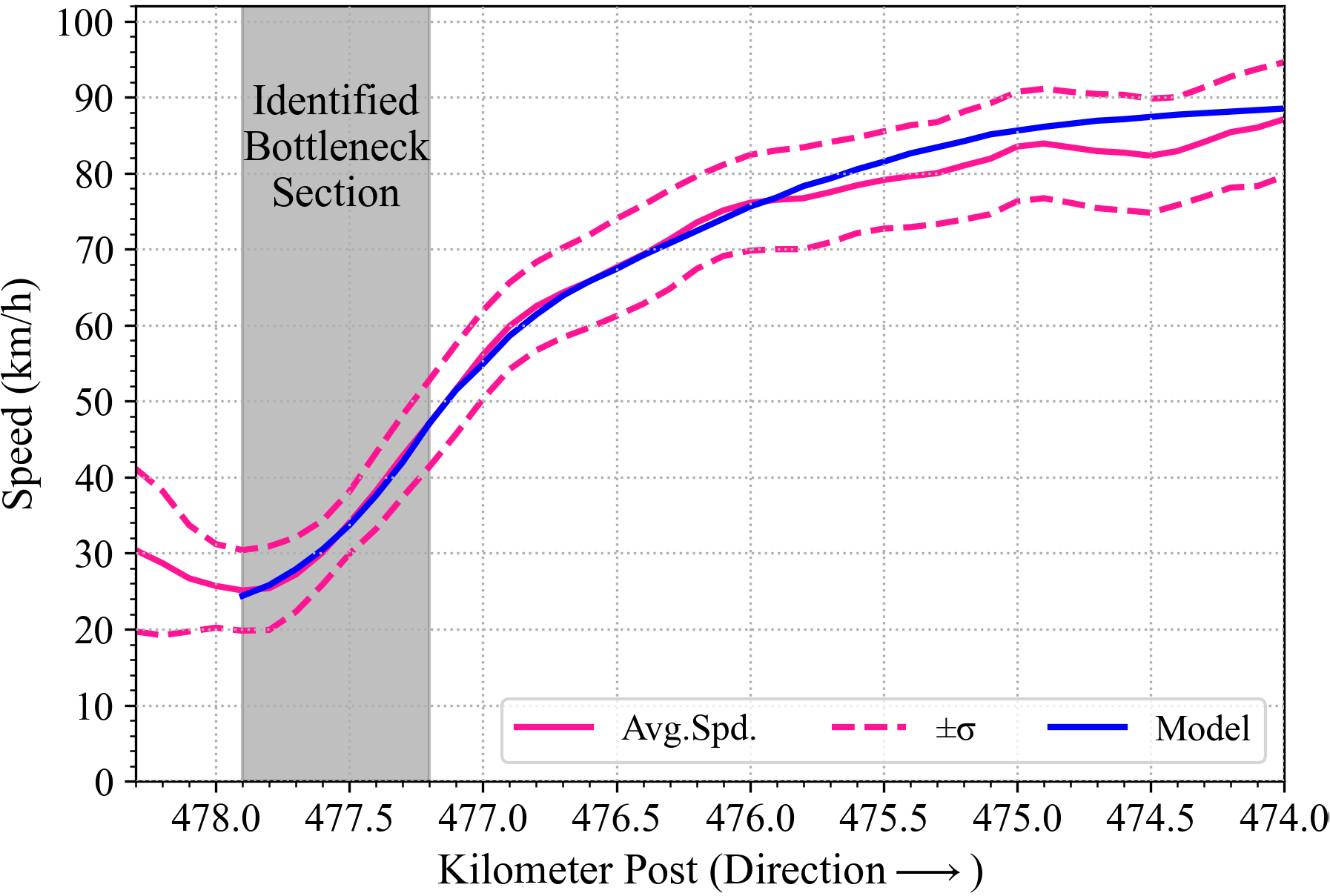}
\caption{Example of model calibration for Semimaru (Mar. 25, 2019)}
\label{Fig4-2}
\end{figure}

\begin{figure}[tbp]
\centering
\includegraphics[height=5.4cm]{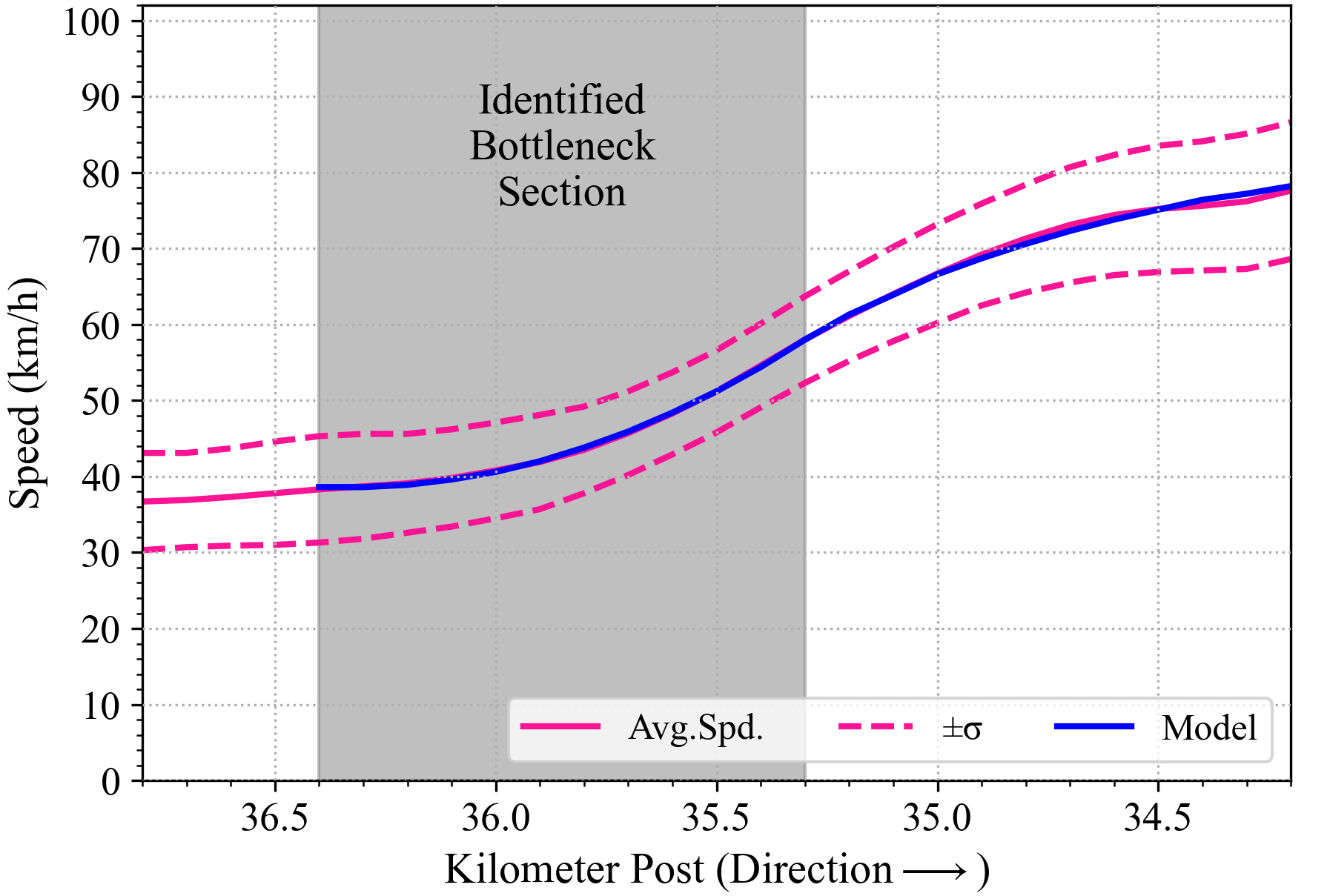}
\caption{Example of model calibration for Takasaka (Nov. 13, 2019)}
\label{Fig4-3}
\end{figure}

\begin{figure}[tbp]
\centering
\includegraphics[height=5.4cm]{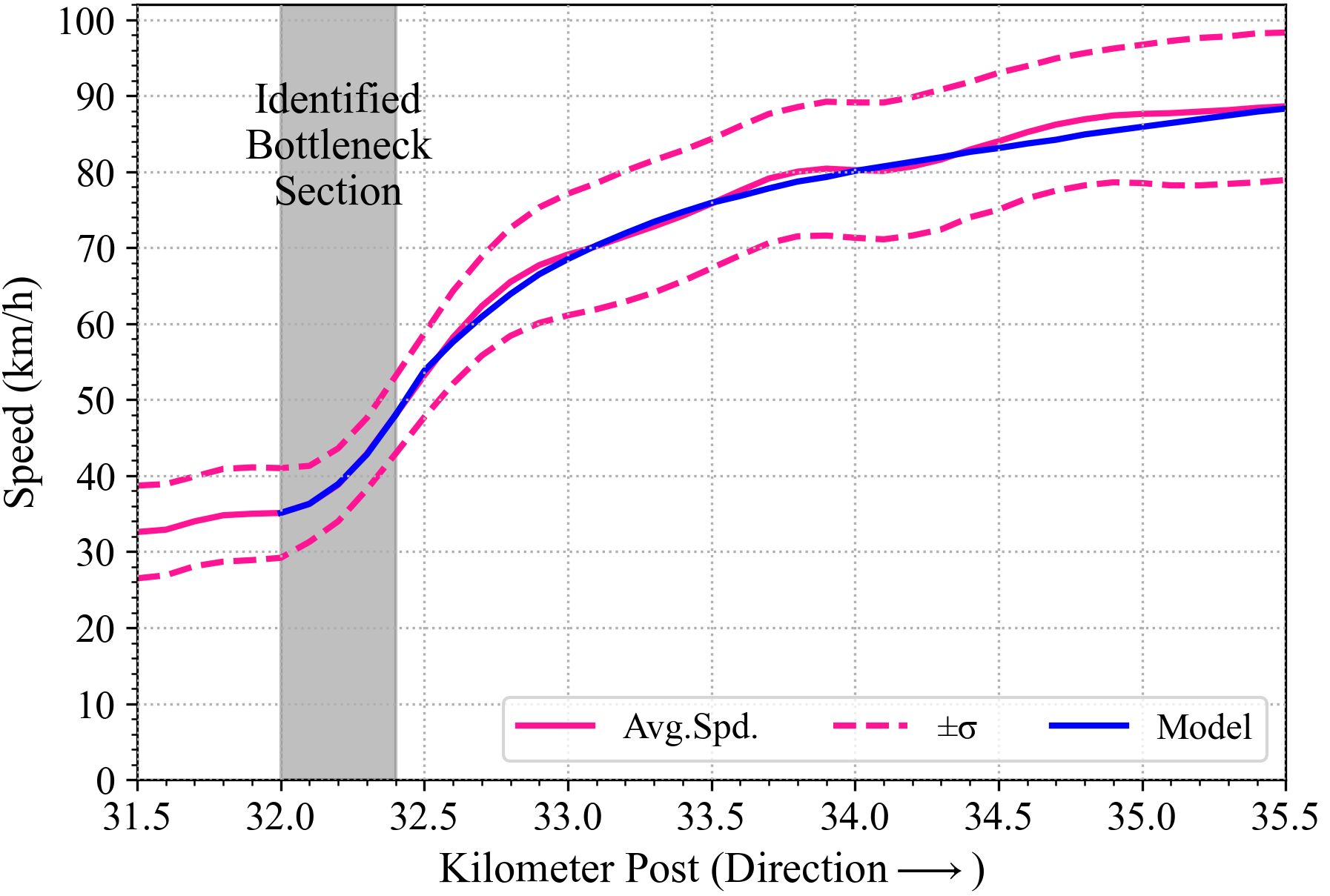}
\caption{Example of model calibration for Komagawa (Sep. 21, 2019)}
\label{Fig4-4}
\end{figure}

For each site, model calibration was performed for 10 congestion events.
A typical speed profile from each section is shown in Figures \ref{Fig4-1} to \ref{Fig4-4}.
The vertical axis represents the speed [km/h], and the horizontal axis represents the location [Kilometer Post]\footnote{{[Kilometer Post] represents the distance in kilometers from the starting point of each expressway line, that is, the location along the expressway.}}, which increases from left to right in the direction of travel. 
The pink solid line indicates the space-mean speed calculated from individual probe data, while the dashed lines denote the standard deviation. 
The blue solid line shows the speed calculated from the calibrated model, and the identified bottleneck section is shaded in gray. 
These figures demonstrate that the calibrated model fits the stationary speeds in the vicinity of the bottleneck {section} very well.

\begin{table*}[tb]
\caption{Estimated parameters}
\label{Tab2}
\center
\begin{tabular}{lcccccccccccc}
\hline
\multicolumn{1}{c}{\multirow{2}{*}{\begin{tabular}[c]{@{}c@{}}Target site\\ \textcolor{black}{(10 samples for each site)}\end{tabular}}} & \multicolumn{2}{c}{\begin{tabular}[c]{@{}c@{}}upstream end of\\ {bottleneck section}\\ {[}Kilometer Post{]}\end{tabular}} & \multicolumn{2}{c}{\begin{tabular}[c]{@{}c@{}}downstream end of \\{bottleneck section}\\ {[}Kilometer Post{]}\end{tabular}} 
& \multicolumn{2}{c}{\begin{tabular}[c]{@{}c@{}}$\tau(0)$\\ {[}s{]}\end{tabular}}
& \multicolumn{2}{c}{\begin{tabular}[c]{@{}c@{}}$\tau(L)$\\ {[}s{]}\end{tabular}}
& \multicolumn{2}{c}{\begin{tabular}[c]{@{}c@{}}$a_0-g\Phi(L)$\\ {[}m/s$^2${]}\end{tabular}}
& \multicolumn{2}{c}{\begin{tabular}[c]{@{}c@{}}$\tau_x(L)$\\ {[}s/km{]}\end{tabular}}\\
\multicolumn{1}{c}{} & Avg. & SD & Avg. & SD & Avg. & SD & Avg. & SD & Avg. & SD & Avg. & SD \\ \hline
Nissin, Tomei & 321.6 & 0.00 & 321.1 & 0.03 & 1.63 & 0.10 & 1.77 & 0.09 & 0.33 & 0.03 & 0.48 & 0.05 \\ \hline
Semimaru, Meishin & 478.0 & 0.04 & 477.2 & 0.00 & 1.56 & 0.06 & 2.03 & 0.06 & 0.38 & 0.02 & 0.59 & 0.02 \\ \hline
Takasaka, Kan-etsu & 36.4 & 0.03 & 35.3 & 0.05 & 2.11 & 0.08 & 2.39 & 0.09 & 0.66 & 0.09 & 0.33 & 0.04 \\ \hline
Komagawa, Kan-etsu & 32.0 & 0.00 & 32.4 & 0.00 & 1.96 & 0.09 & 2.21 & 0.12 & 0.48 & 0.06 & 0.76 & 0.10 \\ \hline
\end{tabular}
\caption{Indicators used for assessing goodness of fit and validation}
\label{Tab3}
\begin{tabular}{lcccccccccccccc}
\hline
\multicolumn{1}{c}{\multirow{2}{*}{\begin{tabular}[c]{@{}c@{}}Target site\\ \textcolor{black}{(10 samples for each site)}\end{tabular}}} & \multicolumn{2}{c}{\begin{tabular}[c]{@{}c@{}}$C(0)$\\ {[}veh/h/lane{]}\end{tabular}} & \multicolumn{2}{c}{\begin{tabular}[c]{@{}c@{}}$C(L)$\\ {[}veh/h/lane{]}\end{tabular}} & \multicolumn{2}{c}{\begin{tabular}[c]{@{}c@{}}BDF\\ {[}veh/h/lane{]}\end{tabular}} & \multicolumn{2}{c}{\begin{tabular}[c]{@{}c@{}}QDF\\ {[}veh/h/lane{]}\end{tabular}} & \multicolumn{2}{c}{\begin{tabular}[c]{@{}c@{}}CDratio\\ {[}\%{]}\end{tabular}} & \multicolumn{2}{c}{\begin{tabular}[c]{@{}c@{}}RMSE of\\ speed\end{tabular}} & \multicolumn{2}{c}{\begin{tabular}[c]{@{}c@{}}BDF / $C(L)$\end{tabular}} \\
\multicolumn{1}{c}{} & Avg. & SD & Avg. & SD & Avg. & SD & Avg. & SD & Avg. & SD & \multicolumn{1}{l}{Avg.} & \multicolumn{1}{l}{SD} & \multicolumn{1}{l}{Avg.} & \multicolumn{1}{l}{SD} \\ \hline
Nissin, Tomei & 1,863 & 96 & 1,740 & 75 & 1,705 & 76 & 1,540 & 60 & 11.5 & 0.9 & 0.77 & 0.28 & 0.98 & 0.02 \\ \hline
Semimaru, Meishin & 1,940 & 62 & 1,552 & 39 & 1,535 & 68 & 1,398 & 32 & 9.9 & 0.3 & 0.66 & 0.24 & 0.99 & 0.04 \\ \hline
Takasaka, Kan-etsu & 1,507 & 50 & 1,351 & 44 & 1,573 & 71 & 1,274 & 40 & 5.7 & 0.4 & 0.98 & 0.50 & 1.16 & 0.05 \\ \hline
Komagawa, Kan-etsu & 1,608 & 64 & 1,446 & 66 & 1,513 & 30 & 1,308 & 63 & 9.5 & 0.5 & 0.81 & 0.32 & 1.05 & 0.05 \\ \hline
\end{tabular}
\end{table*}

Table \ref{Tab2} presents the mean values and standard deviations of the estimated parameters for 10 congestion events in the target sites.
The parameters ``upstream end {of bottleneck section}" and ``downstream end {of bottleneck section}" columns indicate the locations of both ends of the bottleneck section.
{Table \ref{Tab3} summarizes the indicators used for assessing goodness of fit and validation. 
$C(0)$ and $C(L)$ represent the traffic capacities at the respective locations, calculated using the above parameters. 
The CD ratio is defined as $1 - \mathrm{QDF}/C(L)$, representing the reduction in traffic capacity due to CD.
The RMSE of speed indicates the degree of agreement between the calibrated model and the observed speed profile in the 500 [m] downstream of the bottleneck {section}.
The ratio $\mathrm{BDF} / C(L)$ is used as an indicator for validation in the following subsection.} 

From these tables, {we first observe that the RMSE is close to zero in all cases, suggesting that the model accurately reproduces the observed speed profiles.}
Also, the identified bottleneck section is nearly the same across the 10 congestion events. 
Although the values of $\tau(0)$ and $\tau(L)$ differ by sites, their standard deviations across congestion events at each site are also small.
Similar trends are also observed for $a_0 - g\Phi(L)$ and $\tau_x(L)$. 

{In particular, for acceleration-related parameters, we observe a notable improvement in estimation stability compared with \citet{Kai2023}, which analyzed the same sites\footnote{{We found that in \citet{Kai2023} that the number of lanes at Nisshin, which actually has two lanes, was mistakenly recognized as three. As a result, the values of $C(0)$, $C(L)$, BDF, and QDF, which represent flow rates per lane, became two-thirds of their correct values.}}. For example, \citet{Kai2023} shows several cases where $a_{0}-g\Phi(L)$ exceeds 1.0 [m/s$^2$], which is large for the slow speed recovery known to occur when leaving the sag and tunnel bottlenecks. The values of $a_0 - g\Phi(L)$ and $\tau_x(L)$ also vary widely, with coefficients of variation from 20\% to 30\%, indicating instability in parameter estimation. These issues likely stem from limitations in the previous calibration, such as the lack of probe data smoothing and distortion caused by aligning the BA model's free-flow speed with that of the FD. In contrast, this study yields more reasonable and stable parameter estimates, resulting from improvements that address these limitations.}

\begin{figure*}[tb]
\centering
\includegraphics[width=14cm]{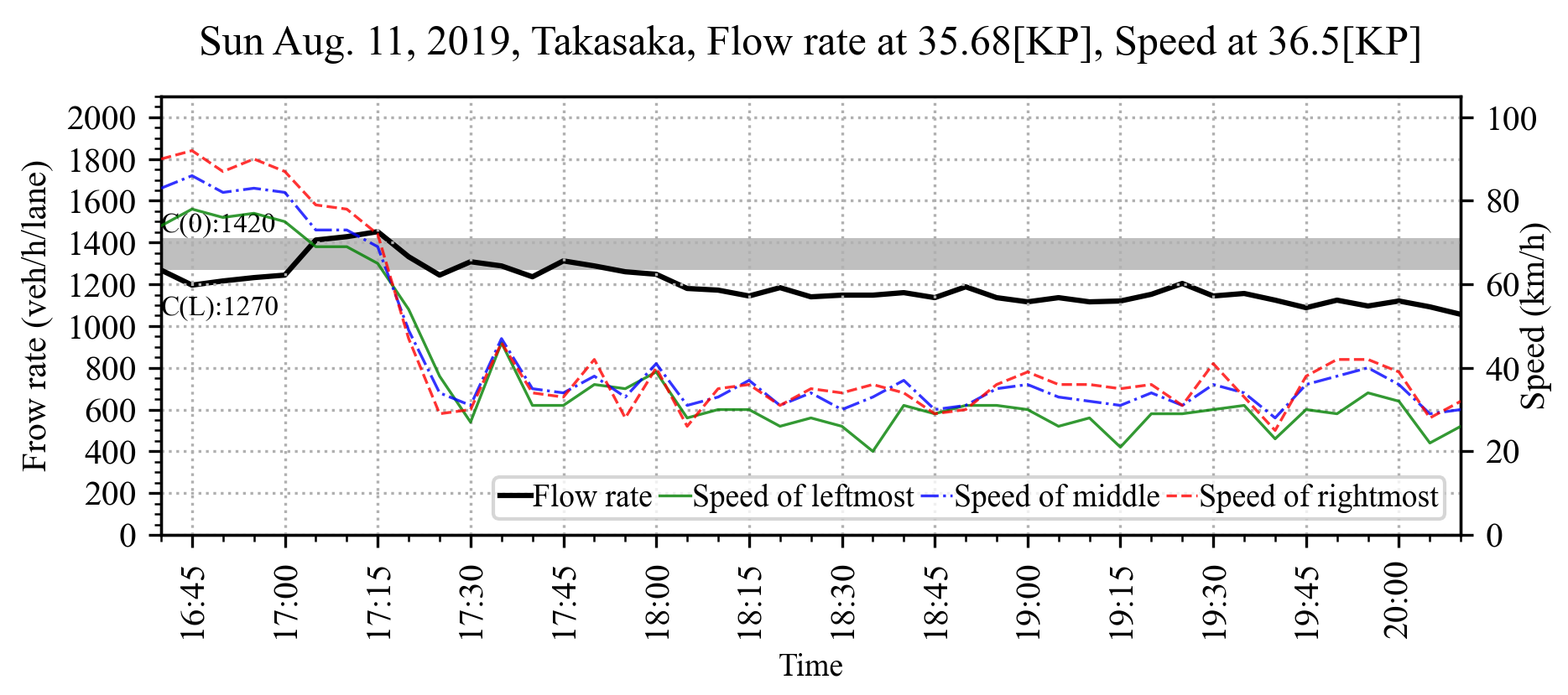}

\caption{An example of variation in speeds between lanes and a further reduction in flow rate during congestion at Takasaka sag}
\label{Fig7}
\end{figure*}

\begin{figure}[tb]
\centering
\includegraphics[height=5.4cm]{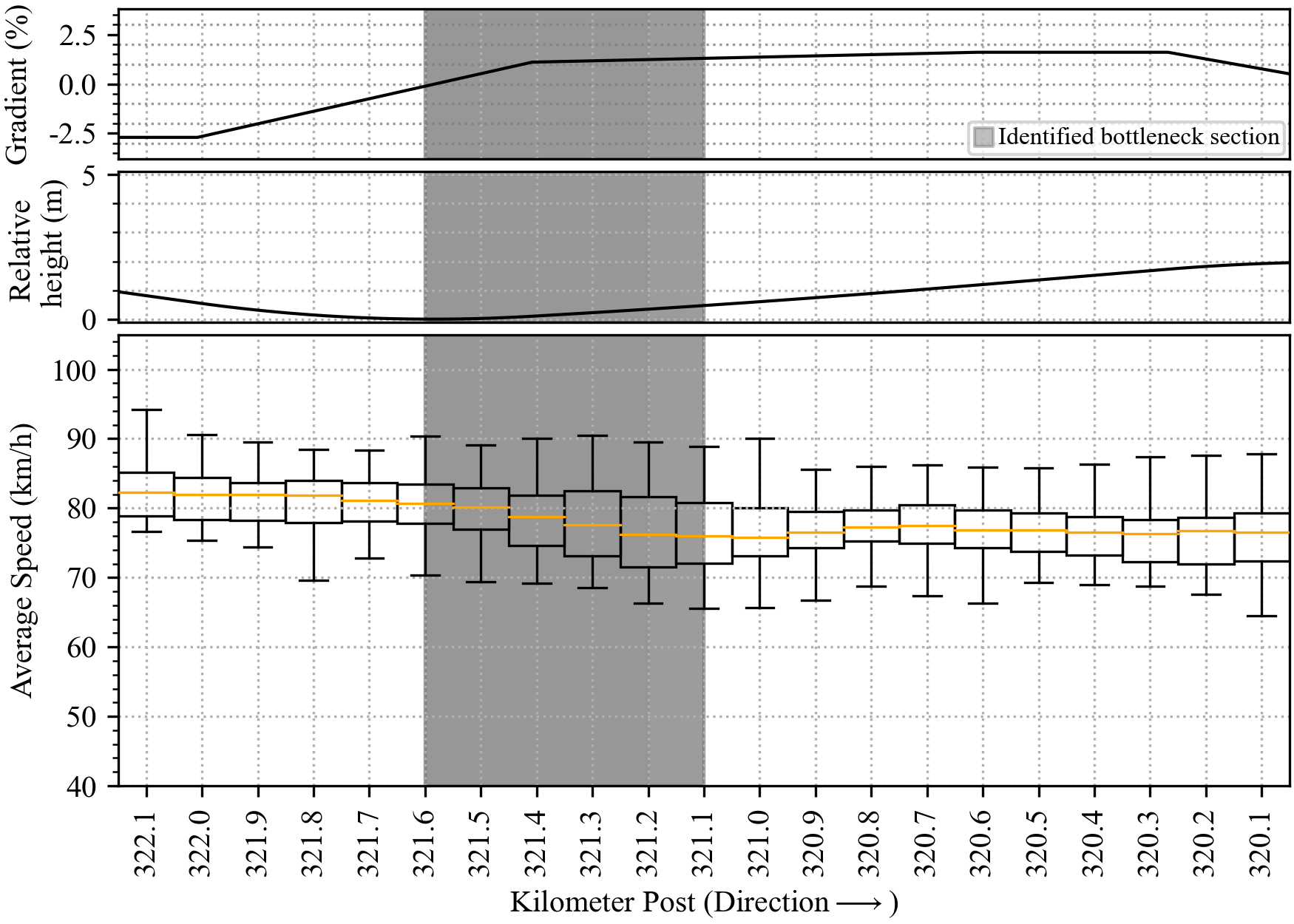}
\caption{Speed distribution during the {15}-minute period before the flow breakdown at Nisshin (10 congestion events)}
\label{Fig5-1}
\end{figure}

\begin{figure}[tb]
\centering
\includegraphics[height=5.4cm]{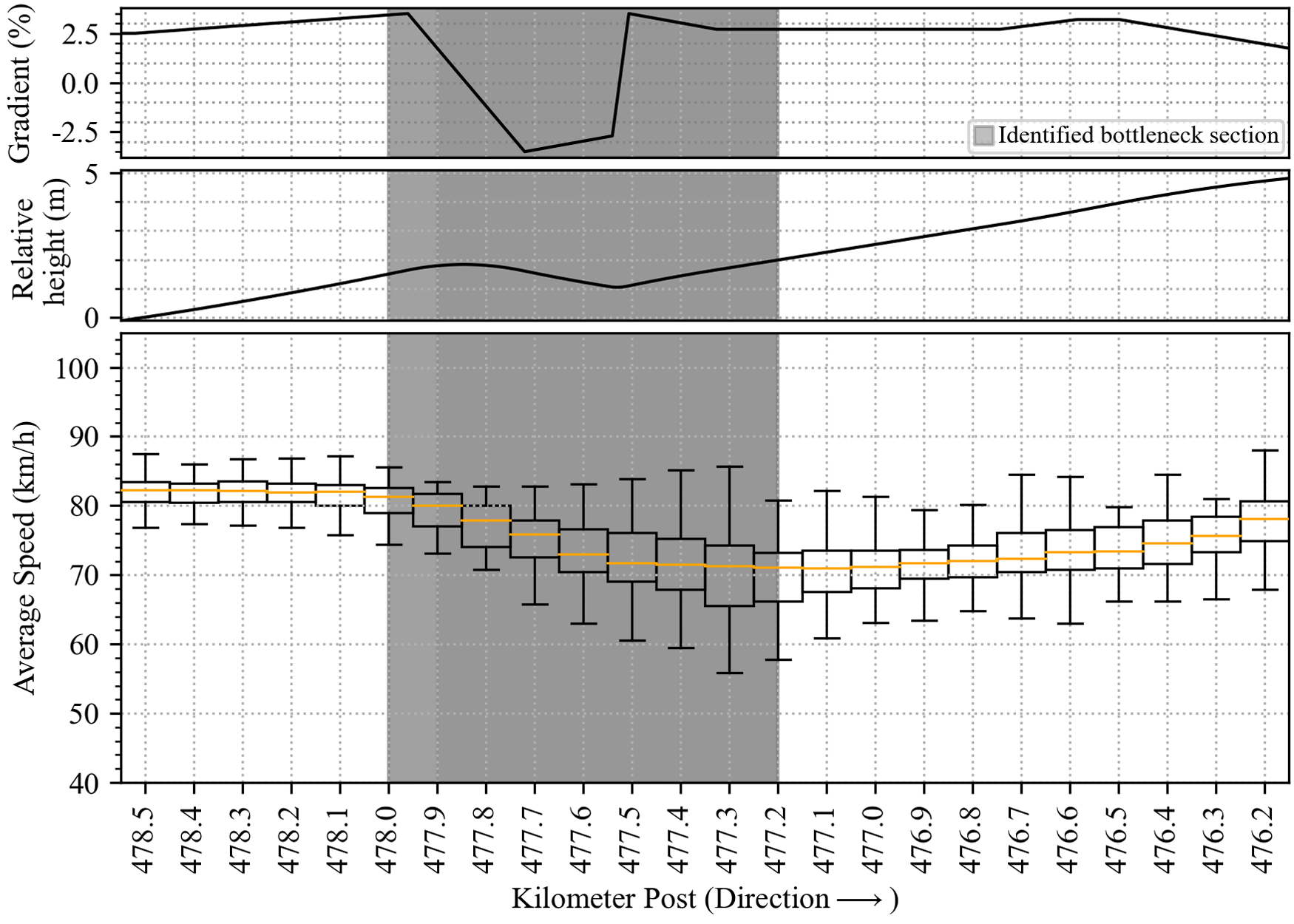}
\caption{Speed distribution during the {15}-minute period before the flow breakdown at Semimaru (10 congestion events)}
\label{Fig5-2}
\end{figure}

\begin{figure}[tb]
\centering
\includegraphics[height=5.4cm]{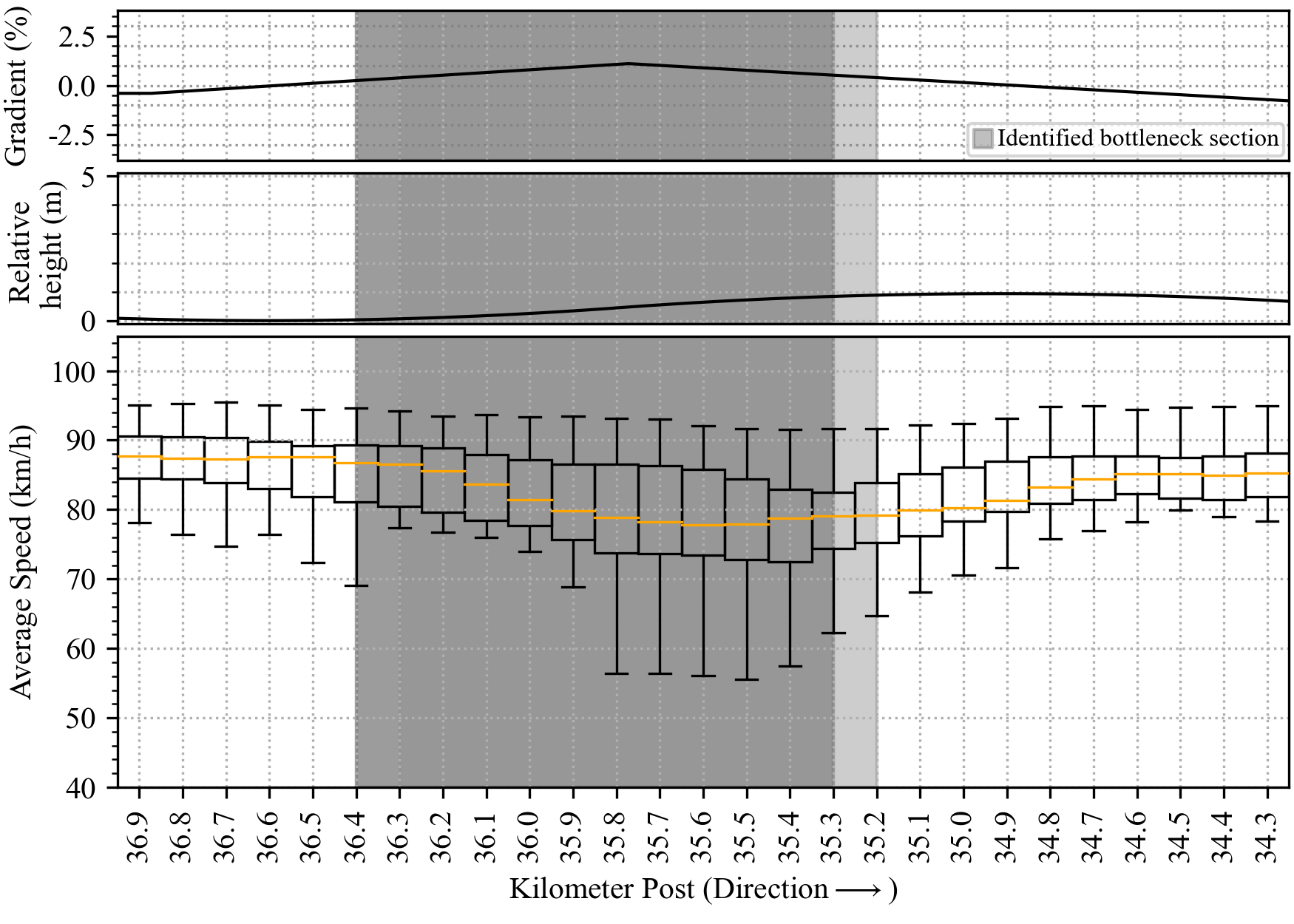}
\caption{Speed distribution during the {15}-minute period before the flow breakdown at Takasaka (10 congestion events)}
\label{Fig5-3}
\end{figure}

\begin{figure}[tb]
\centering
\includegraphics[height=5.4cm]{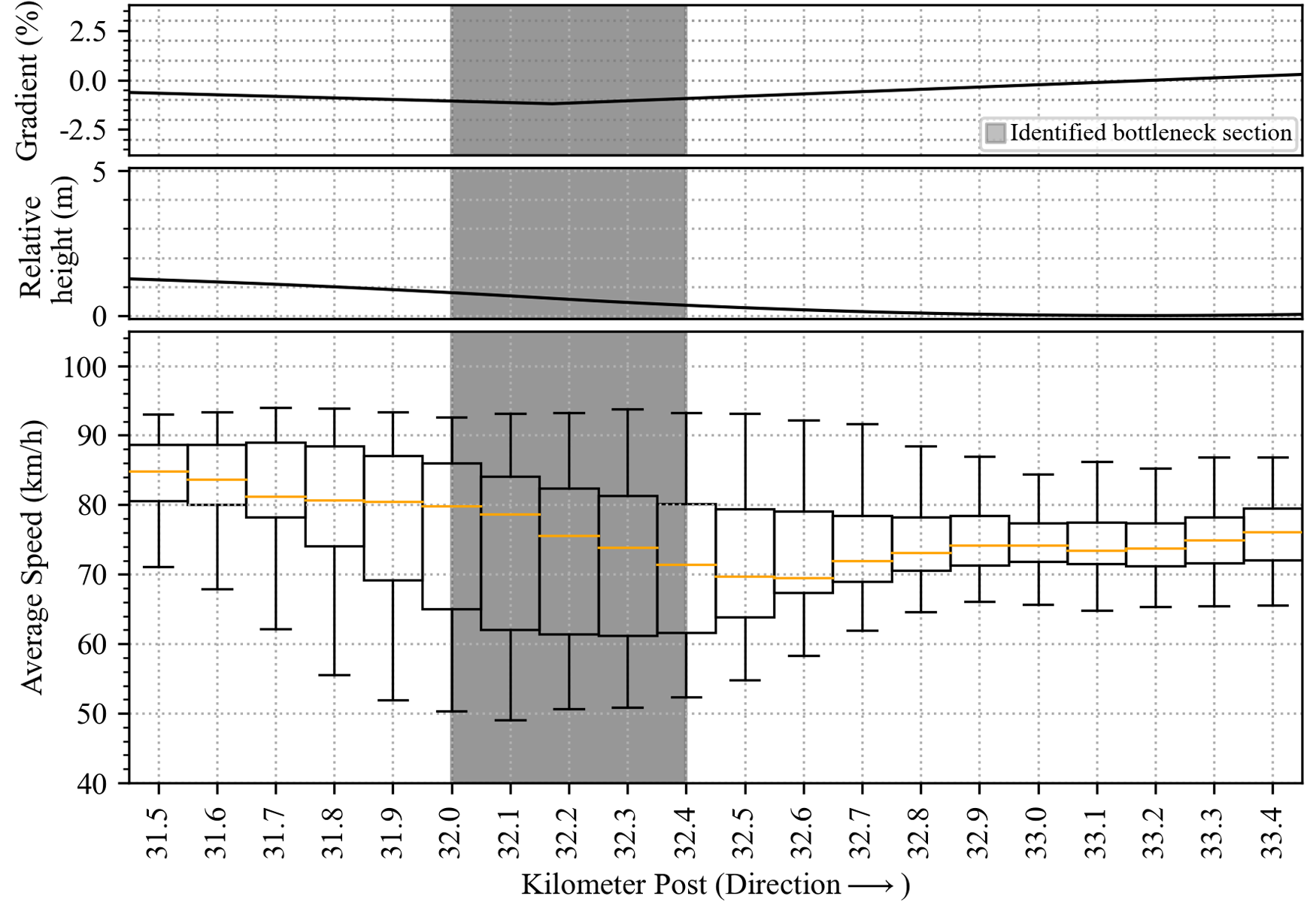}
\caption{Speed distribution during the {15}-minute period before the flow breakdown at Komagawa (10 congestion events)}
\label{Fig5-4}
\end{figure}

\subsection{{Model Validation}}
\label{Ch5.1}

{We first discuss the validation of the estimated capacity.}
Since the values of BDF/$C(L)$ in Table \ref{Tab3} are close to 1 except for Takasaka, this indicates that the capacity estimated by the model is close to the observed traffic flow rate just prior to the flow breakdown, and is therefore reasonable for these sites.
In contrast, the value of BDF/$C(L)$ at Takasaka is somewhat greater than 1. 
 {At Takasaka, although the sag is the main cause of congestion, the nearby SA, which often attracts high demand from drivers seeking rest or services during prolonged congestion, may also influence traffic conditions.}
 More specifically, vehicles entering the SA tend to concentrate in the leftmost lane\footnote{On Japanese expressways, vehicles drive on the left side of the road, so the leftmost lane serves as the lane for vehicles exiting to off-ramps. In addition, the rightmost lane is used as the passing lane.}, resulting in a decrease in speed. 
This may lead to a further decrease in QDF, which results in an underestimation of the capacity $C(L)$.

Such a situation is observed, as demonstrated in Figure~\ref{Fig7}.
This figure shows the traffic flow rate and speed for a congestion event\footnote{This particular congestion event is not included in the calibration results, as the flow breakdown was influenced by an upstream merging section.}, as recorded by the loop detectors near Takasaka. 
The green solid line represents the speed of the leftmost lane, the blue dashed-dotted line represents the speed of the middle lane, and the red dashed line represents the speed of the rightmost lane.
The thick black line represents the 5-minute flow rate. 
For reference, the estimated traffic capacities $C(0)$ and $C(L)$ are also shown ({the gray area
represents the range from $C(0)$ to $C(L)$}). 
As illustrated in this figure, {the 5-minute flow rate, which was around 1400 [veh/h/lane] at the time of flow breakdown, first decreases to about 1250 [veh/h/lane],} and then further declines to below 1200 [veh/h/lane] after 18:00 as the speed difference between the lanes increases during this period.

Thus, even if the model fits well, the estimated traffic capacity may be underestimated when factors other than the CD phenomenon further reduce QDF. {However, this is still meaningful, as it implies physical consistency in the model and its estimation method. If a model gave plausible results under conditions outside its assumptions, it might fail to capture real causal relationships.}

{Next, we turn to the discussion of the validity of the bottleneck location.}
{The upper parts of Figures \ref{Fig5-1} to \ref{Fig5-4} show the longitudinal gradient [\%]. 
The middle parts present the relative elevation profiles within these sections, based on the longitudinal gradient. 
The lower parts of the figures show boxplots of the speed distribution at each location during the {15}-minute period before the flow breakdown, with the orange line indicating the median.}
{For each location, we calculate average speeds at 15, 10, and 5 minutes prior to the flow breakdown for 10 congestion events, yielding a total of 30 samples. The shades of gray in Figures \ref{Fig5-1} to \ref{Fig5-4} indicate how many times each segment was identified as part of the estimated bottleneck section across the ten results. Segments shown in lighter gray were included in the bottleneck section in only a subset of the estimation results.}

{The lower figures show that speed reductions prior to flow breakdown tend to occur near the downstream end of the identified bottleneck {sections}, where capacity is at a minimum.
{This observation is consistent with the description of the model, in which flow breakdown occurs at the downstream end of the bottleneck, while speed recovery during congestion begins at the upstream end. Thus, we can conclude that the identified bottleneck sections are reasonable.}}

{Overall, the two validation results provide evidence that the model is generally valid.} 
{These findings also indicate that the model and calibration method are generally applicable and not limited to the site analyzed in \citet{Wada2022}.}

\subsection{{Application}}
\label{Ch5.2}
Finally, as an application of the calibration results, we examine the relationship between the identified bottleneck sections and changes in longitudinal gradients.
From the top two figures in Figures \ref{Fig5-1} to \ref{Fig5-4}, we can see that the relationship varies across sites. 
At Nisshin (Fig. \ref{Fig5-1}), the bottleneck {section} is identified near the almost sag bottom.
At Semimaru (Fig. \ref{Fig5-2}), the bottleneck {section} is identified from the downhill to the uphill.
At Takasaka (Fig. \ref{Fig5-3}), the bottleneck {section} is identified from the sag bottom to near the crest.
At Komagawa (Fig. \ref{Fig5-4}), the bottleneck {section} is identified within the downhill section where the grade increases.
These observations highlight the importance of accurately identifying bottleneck sections at each sag in order to implement effective countermeasures. 

\section{Conclusion}
\label{Ch6}
This study validated the continuum traffic flow model for sag and tunnel bottlenecks through empirical analysis on several expressway sites and congestion events.
The model provides a reliable means of estimating key bottleneck characteristics, thereby supporting the implementation of effective congestion countermeasures, as discussed in the Introduction.
The results also support the validity of the model, which in turn suggests that sluggish driving behavior induced by the bottleneck is a primary cause of capacity drop. 
{Furthermore, analysis of the calibration results revealed that the relationship between the estimated bottleneck sections and longitudinal gradients varies across sites, highlighting the importance of accurately identifying the bottleneck section at each sag.}

As future work, the calibration method and the estimated bottleneck characteristics could be applied to the design of practical countermeasures.
For instance, \citet{Kimura2024} applied the method to determine the location and message design of LED signs and reported favorable effects.
Another promising direction is to further investigate the relationship between bottleneck locations and points where the road gradient changes.
This line of research could offer important insights into the geometric design of expressways, particularly with respect to longitudinal gradients.
{Although this study focused on empirical validation for sag sections, tunnels are also within the scope of the model, as demonstrated in previous studies \citep{Jin2018, Wadaetal2020}. Validation using tunnel sites remains an important subject for future research.}
Finally, we are also interested in extending the scope of this research to other types of bottlenecks, such as merging sections.
For further details on one example in this direction, see \citet{Hattori2025}.

\section*{Declarations}

\section*{Ethics approval and consent to participate}
Not applicable.
\vspace{-5mm}

\section*{Consent for publication}
Not applicable.
\vspace{-5mm}

\section*{Availability of data and materials}
The data that support the findings of this study are available from the corresponding author upon reasonable request.\\
\vspace{-5mm}

\section*{Competing interests}
The authors declare that they have no competing interests. 

\section*{Funding}
JSPS Grant-in-aid (KAKENHI) \#23K26218.
\vspace{-5mm}

\nopagebreak
\section*{Authors' contributions}
\textbf{SK:} Methodology, Investigation, Software, Formal analysis, Validation, Writing - origianl draft, 
\textbf{RH:} Supervision, Resource,
\textbf{JX:} Data curation, Writing – review \& editing,
\textbf{KW:} Conceptualization, Methodology, Software, Validation, Writing – review \& editing, Supervision, Funding acquisition.\\
\vspace{-5mm}

\section*{Acknowledgements}
The authors express their gratitude to two anonymous referees for their careful reading of the manuscript and useful suggestions. 
This work was partially supported by JSPS Grant-in-aid (KAKENHI) \#23K26218.

%
%

\bibliographystyle{spbasic}      
\bibliography{ijitref.bib}   

%
%

\end{document}